\documentclass{iau}
\usepackage{graphicx}

\title[JD 11.~~Non-Gaussianity in the Planck data] 
{Searching for non-Gaussianity in the Planck data}

\author[Marcelo J. Rebou\c{c}as \& Armando Bernui]   
{Marcelo J. Rebou\c{c}as$^1$
 \and Armando Bernui$^2$}

\affiliation{$^1$Centro Brasileiro de Pesquisas F\'{\i}sicas, \\
Rua Dr.\ Xavier Sigaud 150,
22290-180 Rio de Janeiro -- RJ, Brazil \\ email: {\tt reboucas@cbpf.br}
\\[\affilskip]
$^2$Observat\'orio Nacional\\ Rua General Jos\'e Cristino 77,
20921-400 Rio de Janeiro -- RJ, Brazil \\ email: {\tt bernui@on.br}}

\pubyear{2014}
\volume{306}  
\pagerange{119--126}
\jname{Statistical Challenges in 21$^{st}$ Century Cosmology}
\editors{A.F. Heavens, J.-L. Starck, A. Krone-Martins, eds.}
\begin{document}

\maketitle

\begin{abstract}
The statistical properties of the temperature anisotropies and polarization of the
of cosmic microwave background (CMB) radiation offer a powerful probe of the physics
of the early universe.
In recent works a statistical procedure based upon the calculation of the kurtosis
and skewness of the data in patches of CMB sky-sphere has been proposed and used
to investigate the large-angle deviation from Gaussianity in WMAP maps.
Here  we briefly address the question as to how this analysis of Gaussianity
is modified if the foreground-cleaned Planck maps are considered.
We show that although the foreground-cleaned Planck maps present significant
deviation from Gaussianity of different degrees when a less severe mask is used,
they become consistent with Gaussianity, as detected by our indicators, when masked
with the union mask U73.
\end{abstract}

\firstsection 
\section{Introduction}

The statistical properties of the temperature fluctuations and polarization of
cosmic microwave background (CMB) radiation offer a powerful probe of the physics
of the early universe (\cite[Komatsu 2010]{Komatsu2010}).
In this way, a detection of a significant level of primordial non-Gaussianity (NG)  of local
type ($f_{\rm NL}^{\rm local}\gg 1$) would rule out, for example,  the entire
class of single scalar field models (see, e.g., \cite[Creminelli \& Zaldarriaga 2004]{CZ2004}
and \cite[Komatsu 2010]{Komatsu2010}).

It is conceivable, however,  that no single statistical estimator can be sensitive and
suitable to capture all forms of non-Gaussianity that may be present in the observed CMB data.
Thus,  it is important to test CMB data for non-Gaussianity by using different statistical
indicators.

In a recent paper (\cite[Bernui \& Rebou\c{c}as 2009]{BR2009}) statistical procedure based
upon the calculation of the skewness and kurtosis by taking the values of the CMB temperatures
fluctuations assigned to the pixels inside patches of CMB sky-sphere has been proposed
and used to study deviation from Gaussianity in  foreground-reduced WMAP maps (\cite[Bernui
\& Rebou\c{c}as 2010]{BR2010}) as well as in simulated maps (\cite[Bernui \&
Rebou\c{c}as 2012]{BR2012}).
A pertinent question is how the analysis of Gaussianity made by using WMAP data
is modified if the foreground-cleaned maps released  by the Planck  are considered.
We have addressed this question and here we report partially the results of our analyses
performed with the skewness estimator.
For a comprehensive statistical analysis we refer the readers to our
recent paper (\cite[Bernui \& Rebou\c{c}as 2015]{BR2015}).

\section{Statistical procedure and main results}

Perhaps the simplest test for Gaussianity of a  CMB map can be made by computing  the
skewness, $S$, and kurtosis, $K$ from the whole set of CMB temperature fluctuations values
of a given CMB map. However, one can go a step further and, instead of calculate two numbers,
one can compute $n$ values of the skewness as well as  $n$ values of the kurtosis, and obtain
with directional information on NG, by dividing the CMB sphere $\mathbb{S}^2$ into a number $n$
of uniformly distributed spherical patches of equal area that cover $\mathbb{S}^2 $,
and by calculating the skewness and the kurtosis
\begin{eqnarray}
&&S_j   =  \frac{1}{N_{\rm p} \,\sigma^3_{\!j} } \sum_{i=1}^{N_{\rm p}}
\left(\, T_i\, - \overline{T_j} \,\right)^3 \,, \label{S-Def} \\
&&K_j   =  \frac{1}{N_{\rm p} \,\sigma^4_{\!j} } \sum_{i=1}^{N_{\rm p}}
\left(\,  T_i\, - \overline{T_j} \,\right)^4 - 3 \label{K-Def} \,,
\end{eqnarray}
for each patch $j=1, \ldots ,n$. Here $N_{\rm p}$ is the number of pixels
in the $j^{\,\rm{th}}$ patch, $T_i$ is the temperature at the $i^{\,\rm{th}}$ pixel,
$\overline{T_j}$ is the CMB mean temperature in the $j^{\,\rm{th}}$ patch, and
$\sigma$ is the standard deviation. In this work, we have chosen these patches
to be spherical caps (calottes) with aperture $\gamma = 90^{\circ}$.

The two set of $n$ values  (each) $\{S_j\}$ and  $\{K_j\}$ along with the spherical
coordinates of the center of the patches, $\theta_j,\phi_j$
can then be employed to define two discrete functions on $\mathbb{S}^2$, namely
$S(\theta_i,\phi_i)$ and $K(\theta_i,\phi_i)$ in such way that $S(\theta_j,\phi_j)= S_j$ and
$K(\theta_j,\phi_j)=K_j$ for every $j=1, \ldots ,n$.  These functions give
local measurements of NG as functions of angular coordinates. 
The Mollweide projections of $S(\theta_j,\phi_j)$ and $K(\theta_j,\phi_j)$
are skewness and kurtosis maps, whose power spectra $S_{\ell}$ and $K_{\ell}$
can be used to study large-angle deviation from Gaussianity by
determinig the goodness of fit of these power spectra obtained from the Planck
maps as compared to the mean power spectra calculated from $1\,000$  simulated Gaussian
maps ($\overline{S}^{G}_{\ell}$ and $\overline{K}^{G}_{\ell}$) through a $\chi^2$ analysis.
In this way, for $S_\ell$ obtained from a given Planck
map one has%
\begin{equation}
\chi^2_{S_\ell} = \frac{1}{N-1} \sum_{\ell=1}^{N}\frac{\left({S_\ell}
- {\overline{S}^{G}_{\ell}} \right)^2}{(\,{\sigma_\ell^G}\,)^2}\,,
\label{chi squared}
\end{equation}
where $\overline{S}^{G}_{\ell}$ are the mean multipole values for each $\ell$
mode, $(\sigma_{\ell}^{G})^2 $  is the variance computed from $1\,000$ Gaussian
maps, and $N$ is the highest multipole taken in the analysis of NG.

Clearly a similar expression and reasoning can be used for $K_\ell$.
In what follows, however, for the sake of brevity we will only briefly report
the results of our analysis related to the skewness. For a comprehensive statistical
analysis see our recent paper (\cite[Bernui \& Rebou\c{c}as 2015]{BR2015}).  

\begin{figure*}[h!] 
\begin{center}
\includegraphics[width=6cm,height=4.0cm]{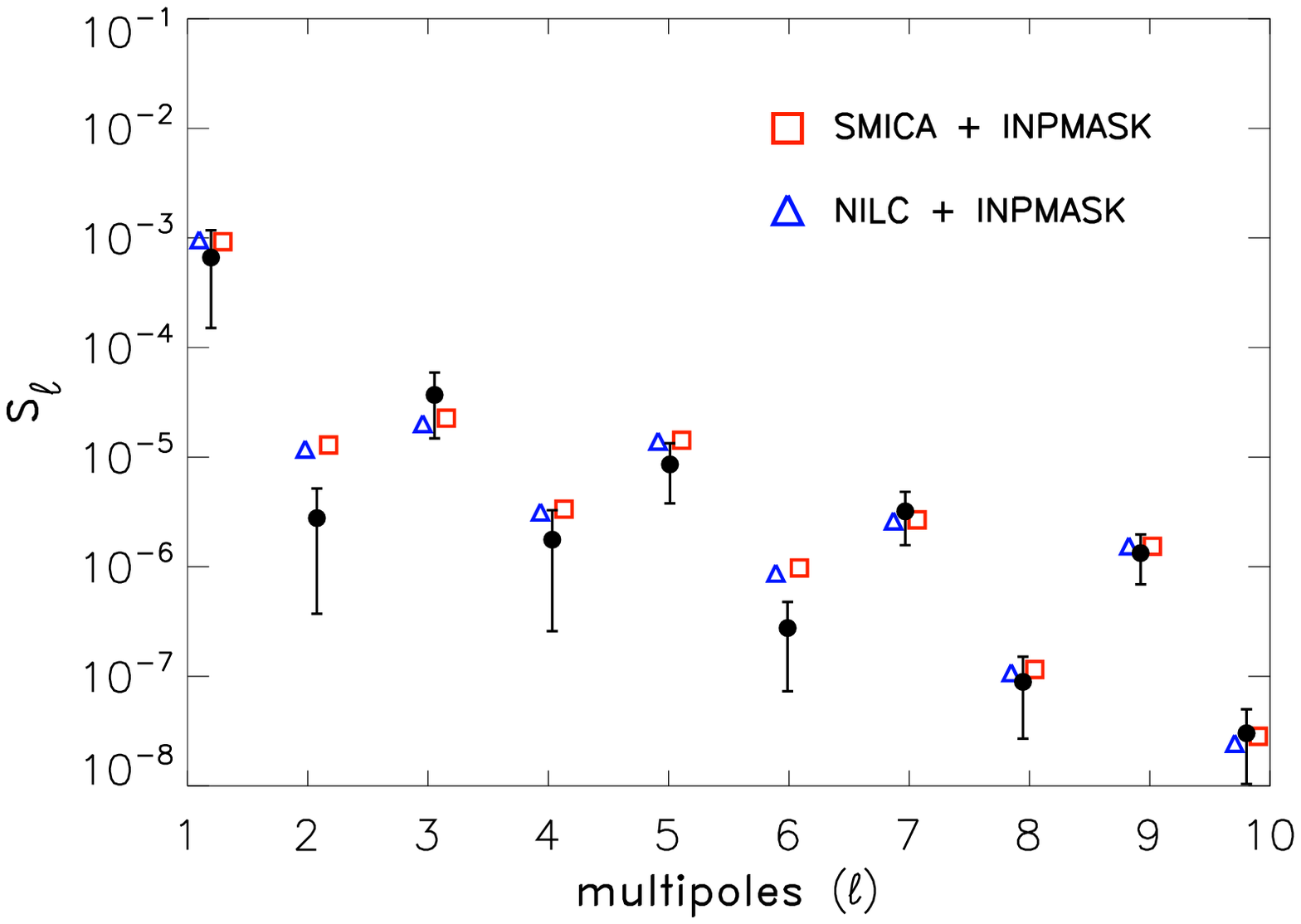}
\hspace{3mm}
\includegraphics[width=6cm,height=4.0cm]{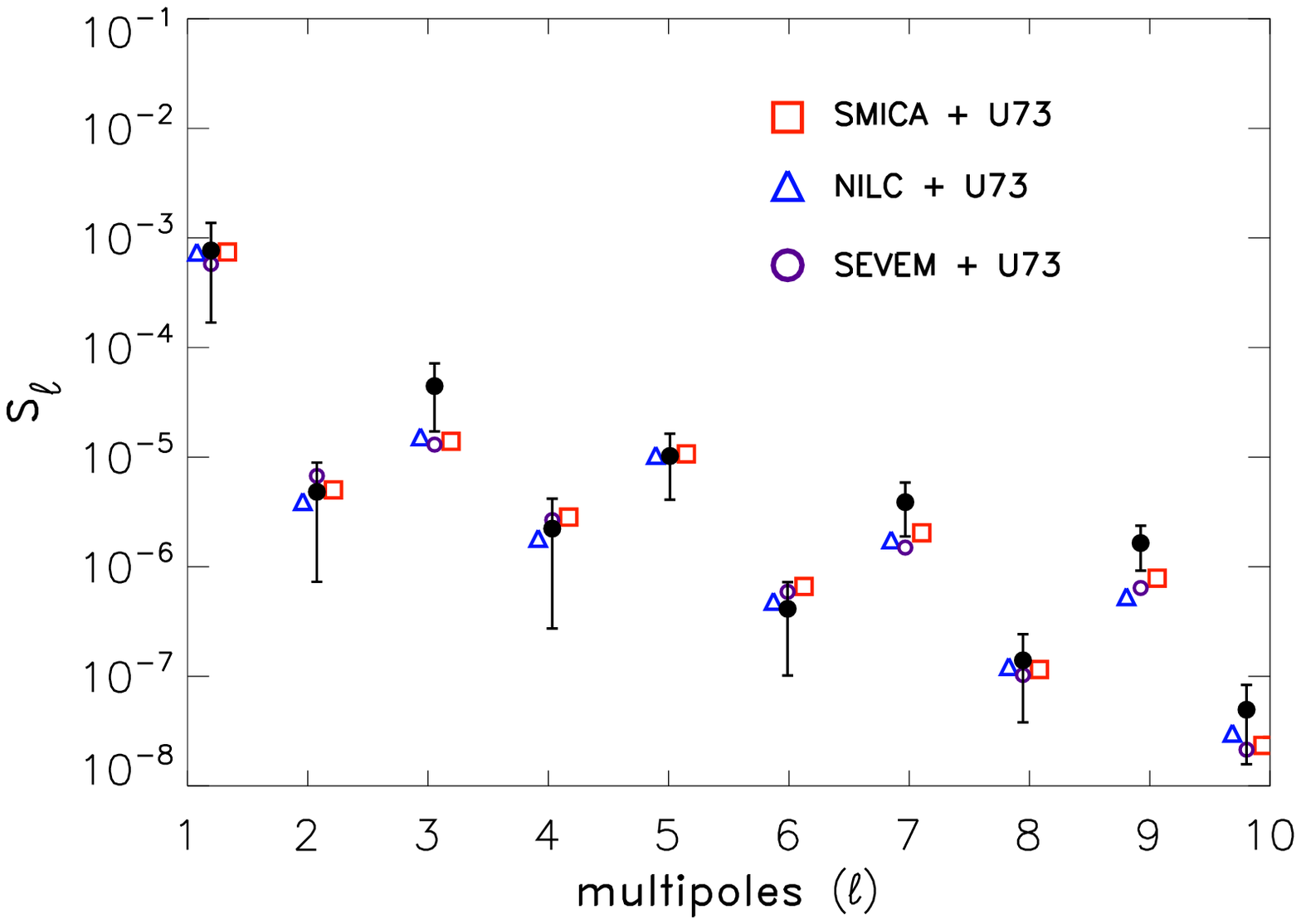}
\caption{Low $\ell$ power spectra $S_{\ell}$ calculated from {\sc smica} and
{\sc nilc} Planck maps equipped with {\sc inpmask} (left panel) and
with U73 mask. We note that since there is no available {\sc inpmask} for the
{\sc sevem} map we have not included this map in the analysis with the {\sc inpmask}.
Tiny horizontal shifts were used to avoid overlaps of symbols.}
\label{Fig1}
\end{center}
\end{figure*}

Figure~\ref{Fig1} shows the power spectra $S_{\ell}$  calculated from
{\sc smica} and {\sc nilc}  maps masked with {\sc inpmask} (left panel).
The right panel of this figure shows the power spectra $S_{\ell}$
computed from {\sc smica}, {\sc nilc} and {\sc sevem} maps with the
U73 mask. This figure also contains the points of the averaged
power spectra $\overline{S}^{G}_{\ell}$  calculated
from $1\,000$ Gaussian simulated CMB maps and the
$1\sigma$ error bars.
To the extent that some of power spectra values $S_{\ell}$ fall off
the $1\sigma$ error bars centered at $\overline{S}^{G}_{\ell}$ value,
the left panel of this figure indicates departure  from Gaussianity in
both  {\sc smica} and {\sc nilc} maps when masked  with {\sc inpmask}.
However, the right panel indicates that departure disappear when the
more severe U73 mask is used.

The above comparison of the power spectra by using Fig.~\ref{Fig1}
is useful as a qualitative indication of NG of Planck maps with different masks.
However, to have a quantitative overall assessment of large-angle deviation from
Gaussianity we have used the  power spectra $S_\ell$ (calculated from the
Planck maps) to carry out the above-mentioned  $\chi^2$ analysis to determine the
goodness of fit of $S_\ell$ computed from the Planck maps as compared to the
mean power spectra $\overline{S}^{G}_{\ell}$.
Table~\ref{Table1} makes clear that although with
different $\chi^2$--~probabilities the {\sc smica}, {\sc nilc}
and {\sc sevem} masked with {\sc inpmask} exhibit small level of NG,
but when the union mask U73 is used these maps are consistent with
Gaussianity as detected by our indicator $S$, in agreement with the
results found by the Planck team (\cite[Ade et al. 2013]{Ade_etal2013}).

\vspace{-3mm}
\begin{table}[!hbt]
\begin{center}
\begin{tabular}{lc} 
\hline 
Map \ \& \ Mask \ & $\chi^2_{_{S_\ell}}\!\!$--~probability \\
\hline
{\sc smica}--{\sc inpmask}  &  $1.00\!\times\!10^{-4}$ \\
{\sc nilc}--{\sc inpmask}   &  $1.80\!\times\!10^{-3}$ \\
{\sc smica}--U73            &  $8.43\!\times\!10^{-1}$ \\
{\sc nilc}--U73             &  $8.25\!\times\!10^{-1}$ \\
{\sc sevem}--U73            &  $7.29\!\times\!10^{-1}$  \\
\hline 
\end{tabular}
\end{center}
\caption{Results of the $\chi^2$--~probability test  to determine the goodness of fit
for $S_{\ell}$ multipole values, calculated from the {\sc smica}, {\sc nilc} and {\sc sevem}
with {\sc inpmask} and U73 masks, as compared to the mean  power spectra $\overline{S}^{G}_{\ell}$
obtained  from $1\,000$ simulated Gaussian maps.}  \label{Table1}
\end{table}

\begin{acknowledgments}
M.J. Rebou\c{c}as acknowledges the support of FAPERJ under a CNE E-26/102.328/2013 grant.
M.J.R. and A.B. thank the CNPq for the grants under which this work was carried out.
Some of the results were derived using the HEALPix
package (\cite[Gorski et al. 2005]{Gorski_etal2005}).
\end{acknowledgments}

\end{document}